 \documentclass[final,5p,times,twocolumn,numbers]{elsarticle}

\usepackage{amssymb}
\usepackage{lineno}
\usepackage{hyperref}
\usepackage{listings}
\begin{document}
%\linenumbers

\begin{frontmatter}

%% Title, authors and addresses

%% use the tnoteref command within \title for footnotes;
%% use the tnotetext command for theassociated footnote;
%% use the fnref command within \author or \affiliation for footnotes;
%% use the fntext command for theassociated footnote;
%% use the corref command within \author for corresponding author footnotes;
%% use the cortext command for theassociated footnote;
%% use the ead command for the email address,
%% and the form \ead[url] for the home page:
%% \title{Title\tnoteref{label1}}
%% \tnotetext[label1]{}
%% \author{Name\corref{cor1}\fnref{label2}}
%% \ead{email address}
%% \ead[url]{home page}
%% \fntext[label2]{}
%% \cortext[cor1]{}
%% \affiliation{organization={},
%%            addressline={}, 
%%            city={},
%%            postcode={}, 
%%            state={},
%%            country={}}
%% \fntext[label3]{}

\title{CEDVR: A virtual reality event display for CMS}

\author[first]{Andrew Su}
\author[first]{Nathan Joshua}
\author[first]{Ethan Hsu}
\author[first]{Joseph Malham}
\author[second]{Muhammad Ansar Iqbal}
\author[first]{Jay Hauser}
\affiliation[first]{organization={University of California, Los Angeles},
            addressline={Department of Physics and Astronomy}, 
            country={USA}}
\affiliation[second]{organization={University of Wisconsin-Madison},
            addressline={Department of Physics}, 
            country={USA}}

\begin{abstract}

 Collision event displays are widely used in high-energy physics (HEP) experiments, enabling researchers to visualize and debug raw and reconstructed event data, and to communicate aspects of individual events to broader audiences. Such event displays are typically viewed on a conventional two-dimensional computer screen. To enhance immersion and interactivity with CMS data, we have developed a virtual reality (VR) software package named CEDVR (for CMS Event Display in Virtual Reality) that uses widely-available VR headsets and hand controllers. This application provides an immersive detector-coordinate view and a novel, complementary LEGO-style ($\eta,\phi$) view. CEDVR provides easy access to physics object properties and object filtering, and incorporates novel VR features for navigation and object selection. Compatible with consumer-grade Quest headsets, CEDVR is a useful visualization tool for CMS physicists, and contains built-in sample events that support educational and public outreach initiatives.

\end{abstract}

%%Graphical abstract
%\begin{graphicalabstract}
%\includegraphics{grabs}
%\end{graphicalabstract}

%%Research highlights
%\begin{highlights}
%\item Research highlight 1
%\item Research highlight 2
%\end{highlights}

\begin{keyword}
%% keywords here, in the form: keyword \sep keyword, up to a maximum of 6 keywords
event display \sep virtual reality \sep CMS \sep immersive visualization \sep particle physics software \sep interactive event visualization \sep scientific visualization

%% PACS codes here, in the form: \PACS code \sep code

%% MSC codes here, in the form: \MSC code \sep code
%% or \MSC[2008] code \sep code (2000 is the default)

\end{keyword}

\end{frontmatter}

%\tableofcontents

%% \linenumbers

%% main text

\section{Introduction}
\label{introduction}

Particle physics event displays are an important aspect of modern high-energy physics (HEP) experiments, as they provide intuitive spatial representations of raw data and reconstructed objects within the detector. Event displays can facilitate a better understanding of data, as well as support outreach and educational efforts. For the past several decades, event displays have mostly been viewed on 2D computer screens, or reproduced in print form. However, virtual reality (VR) technology offers several advantages for viewing particle physics events: i) viewing in all directions, often referred to as "360-degree viewing", ii) binocular vision so that the user can easily and intuitively appreciate structure in the third dimension, and iii) motion of the user's point of view, either by physical movement or, for instance, by joystick or other controls on handsets.

VR displays have been developed for several previous particle physics experiments. NeutrinoKAVE~\cite{NeutrinoKAVE2014} visualized neutrino events in Super-K and the future Hyper-K, but required a six-wall projection system and stereoscopic glasses.
ATLASrift~\cite{ATLASrift2015}, developed around 2015 and subsequently extended~\cite{ATLASrift2019}, visualized LHC events in the ATLAS detector using Oculus Rift and other PC-tethered headsets.
Belle2VR~\cite{Belle2VR2020} visualized simulated $e^+e^-$ events in the Belle II detector, including detailed GEANT4 history, and supported fully untethered operation on devices such as the Oculus Go. JUNO VR~\cite{JUNOVR2026} visualizes reactor antineutrino and background interactions in the JUNO neutrino detector, using untethered Meta Quest headsets.

The need for effective visualization is especially relevant for large, complex particle detectors such as the Compact Muon Solenoid (CMS)~\cite{CMS_Experiment_2008} at the Large Hadron Collider (LHC)~\cite{LHCmachine}, the main focus of our work. The scientific goals of the CMS experiment include studies of the Higgs particle, precision studies of the Standard Model, and searches for new physics beyond the Standard Model. These goals rely on the accurate reconstruction and interpretation of a wide range of physics objects produced in proton-proton collisions, including charged-particle tracks, electromagnetic and hadronic energy deposits, and higher level objects such as leptons, photons, jets and missing transverse momentum.

Several CMS-specific event displays have been developed, including Fireworks \cite{Kovalskyi2010Fireworks} and iSpy WebGL \cite{McCauley2017,McCauley2023iSpyWebGL}, hereafter referred to as iSpy. Both of these are fundamentally screen-based applications which rely on either two-dimensional projections or three-dimensional views that are navigated using a mouse and keyboard. Fireworks was designed as a physics analysis tool with its user interface and features geared towards physics analysis workflows including object filtering, multi-view projections, and energy plots. In contrast, iSpy was designed with both analysis and outreach in mind. Its design philosophy focuses on ease of use and installation with fewer analysis-oriented features. While both applications have downloadable executables, iSpy offers a browser-based event display as well that makes it highly accessible.

\begin{figure}[ht!]
    \centering
    \includegraphics[width=1\linewidth]{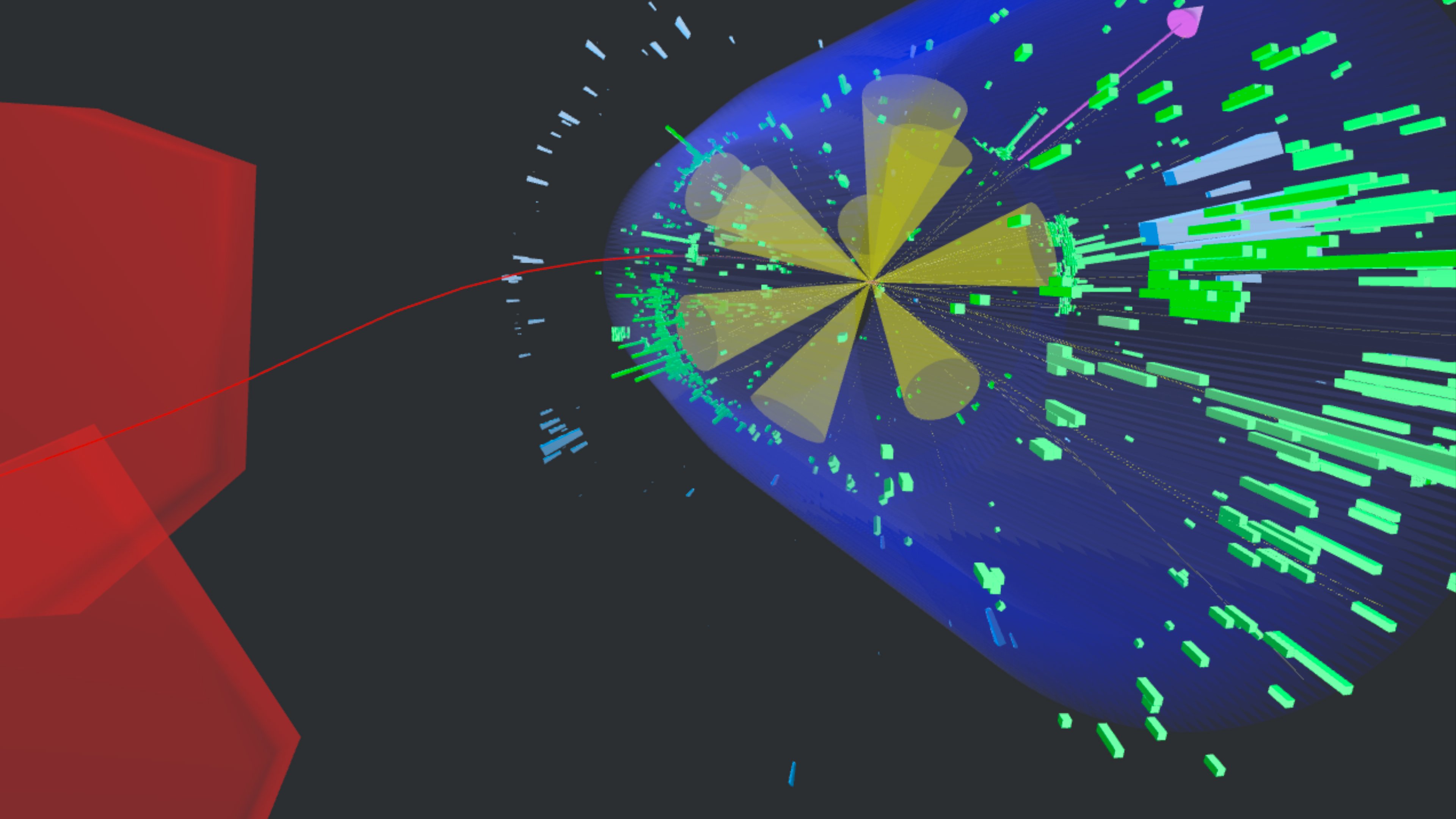}
    \caption{A screen capture from the CEDVR primary view using an event from CMS Open Data \cite{CMSOpenDataJetHTRun2012C}, showing jet cones and charged tracks in yellow, calorimeter energy deposits in green and blue, and a muon reaching outer muon chambers in red.}
    \label{fig:DetectorRefImage}
\end{figure}

In this paper, we present CEDVR, a toolkit for VR displays of CMS events that has been developed to complement the screen-based Fireworks and iSpy displays.  CEDVR offers significant advantages when viewing complex detector geometry and events with dense topologies. 
Our application is compatible with consumer-grade VR headsets, such as the Meta Quest series~\cite{MetaQuest}. In CEDVR, users navigate three-dimensional virtual environments such as that seen in Figure~\ref{fig:DetectorRefImage} by physically moving within their tracked play space, or by using handheld joystick controls. 
Viewing perspective can likewise be adjusted either by natural head movement or through joystick input. The detector geometry and event data surround the user in all directions, which creates a spatially immersive experience. 

The design philosophy of this work was to combine ease of use and installation with a functional physics analysis toolbox in order to create an accessible and powerful application that can be used by both physicists and educators. CEDVR integrates with existing CMS workflows, utilizing the same file format and analyzer code as iSpy.

CEDVR offers a rich feature set, including several novel capabilities: A powerful (up to 60x) zoom capability allows inspection of primary and secondary vertices down to the micron-level spatial resolution of the inner tracker.
An alternate ``LEGO" view generates a two-dimensional histogram of calorimeter energy deposits in $(\eta, \phi)$ space, with overlays showing the directions and energies or momenta of other reconstructed physics objects in the event. By using the LEGO view in conjunction with the regular event display and optional object selection cuts, users gain access to a powerful framework for rapid inspection of event topology.

CEDVR was designed for a mainstream VR setup consisting of a headset and two handheld controllers containing joysticks for navigation and buttons for other controls. The application running in the headset is packaged as an APK for Meta Quest 2/3/3S headsets. To develop the CEDVR headset application, we chose the Unity software platform~\cite{UnityEngine}. Unity is a widely-used engine for developing 2D and 3D applications such as video games and other simulations. Integration with the Meta Quest headsets was achieved using the Meta Software Development Kit (SDK), which supports user mobility and programming of controllers to accommodate a wide variety of user input actions.

\section{Data Pipeline Architecture}
\label{architecture}

\begin{figure}[ht!]
    \centering
\includegraphics[width=1.0\linewidth]{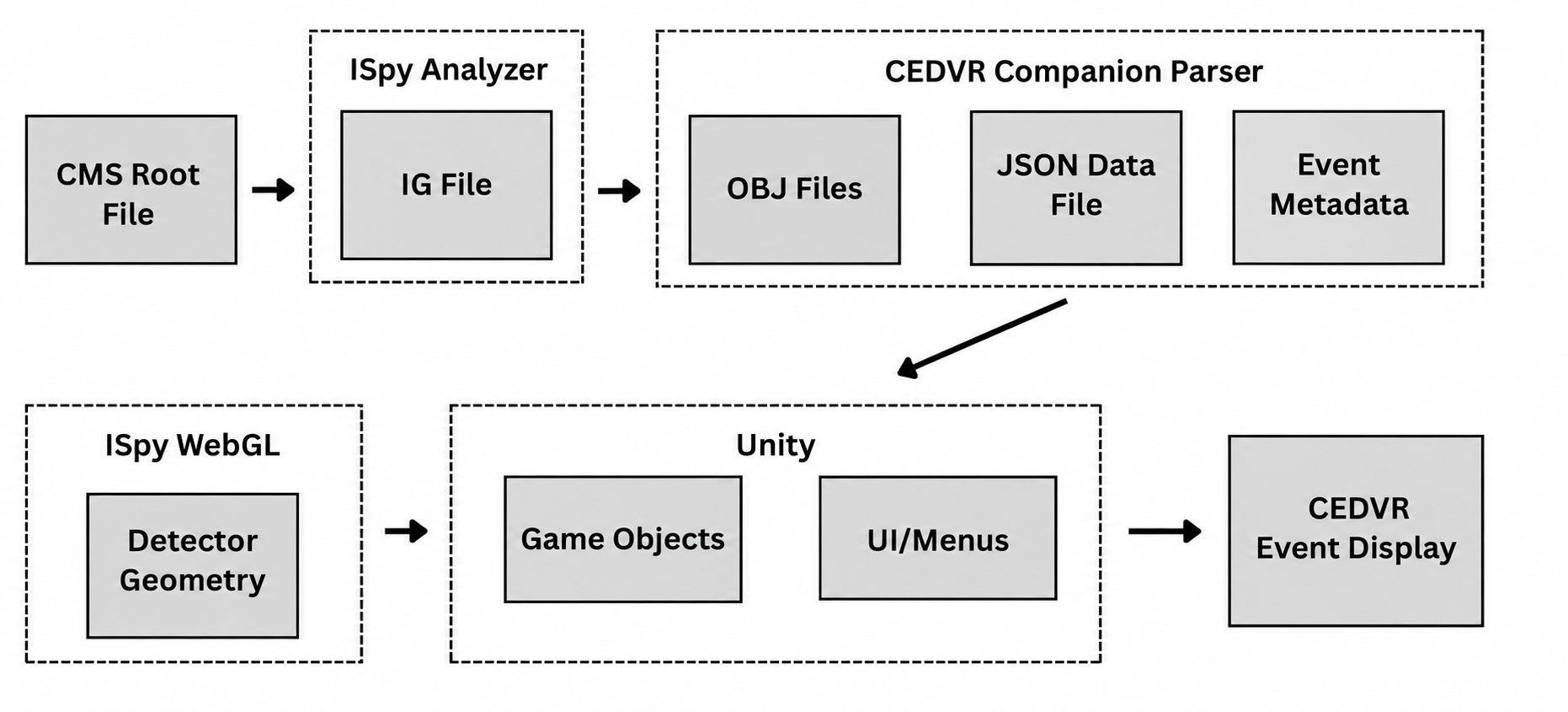}
    \caption{Software framework and Data Pipeline Architecture in CEDVR.}
    \label{fig:CEDVRpipeline}
\end{figure}

The CEDVR data pipeline transforms CMS event data and detector geometry files into graphical objects that can be viewed and manipulated interactively within the VR headset. Steps in the CEDVR data pipeline, shown in Figure~\ref{fig:CEDVRpipeline}, are described in the following subsections.

\subsection{OBJ Format}
\label{objformat}

The CEDVR Companion Parser generates three-dimensional geometries in the form of OBJ files, a widely used geometry definition format\cite{WavefrontOBJ}. Each file contains indexed vertices representing points in three-dimensional space and a set of faces and lines that reference those indices to construct surfaces and shapes. 

\subsection{Detector Geometry}
\label{geometry}

CMS~\cite{CMS_Experiment_2008} is a large general-purpose particle experiment, built for exploring the physics of high-energy proton-proton collisions at the CERN Large Hadron Collider~\cite{LHCmachine}. The CMS detector consists of inner layers of silicon pixel and strip tracking detectors that measure the trajectories of charged particles in a $3.8$~T magnetic field, surrounded by a crystal electromagnetic calorimeter (ECAL), then a brass/scintillator hadronic calorimeter (HCAL). Outside of the HCAL, multiple layers of muon detectors are embedded in the return yoke of the central solenoid coil.

The CMS detector geometry was exported from the iSpy application~\cite{McCauley2017} in the OBJ format, then imported into Unity. The geometry was then scaled to human-scale dimensions suitable for movement within the virtual environment. The detector is represented by thin surfaces rather than closed solids, and the CEDVR user can view these surfaces from any direction, including from within the detector.

\subsection{iSpy Analyzer}
\label{analyzer}

Generally, data from the CMS experiment is stored in the .root format in tiers of complexity. The information created by event reconstruction software is stored in RECO format. Compressed formats such as AOD (Analysis Object Data) and MiniAOD~\cite{MiniAOD} are optimized for general-purpose data analysis, but not for graphics purposes. Typical event sizes are of order several MB for RECO, around 400 kB for AOD, and around 40 kB for miniAOD.
The iSpy package includes analyzers that are implemented using modules from the CMS software framework (CMSSW) \cite{alverson2012ispy}, and produce ig files. The procedure for setting up and executing these analyzers is documented on the iSpy analyzers GitHub repository\cite{iSpyAnalyzers}.
Example configuration files for MiniAOD or AOD tiers are also provided in the iSpy GitHub repository. In some cases, it may be necessary to explicitly include additional CMSSW collections of interest if they are not present in the example configuration file.

\subsection {CEDVR Companion Parser}
\label{parser}

Once an ig file has been obtained, the user must use the CEDVR Companion Parser in order to process and upload data files to the headset's event display. The parser is run from the command line, with the path to the ig file provided as an argument. The processing pipeline proceeds as follows:
\begin{enumerate}
    \item The ig file archive is extracted into a temporary workspace on the user's computer. 
    \item The specific events are selected via user input. From each event, the individual data types (tracks, jets, muon chambers etc.) are parsed and stored into custom structs.
    \item From this parsed information, OBJ files are generated based on the data type. For instance, we represent particle tracks as thin ribbons using cubic Bézier curves and muon chambers as rectangular prisms. These are based on methods implemented in iSpy~\cite{McCauley2017}.
    \item After generating OBJ files for each data type, the total parsed data for the event is compiled and written into a JSON file for use in the event display.
    \item Using the Android Debug Bridge (ADB)—a command-line tool for communicating with and transferring files to Android devices—the processed OBJ and JSON files from each event are uploaded directly to the VR headset.
\end{enumerate}
An example CEDVR Companion Parser session is shown below:

\begin{lstlisting}[
    breaklines=true,
    basicstyle=\ttfamily\small,
    columns=fullflexible
]
$ ./"CEDVR Companion" /path/to/input/MyData.ig c # Running using 'c' select range flag
0) /path/to/temp/Extraction/Events/Run_1 # Available runs          
Enter ID # of desired path:
0   # User input                                
#Available Events
0) /path/to/temp/Extraction/Events/Run_1/Event_1
1) /path/to/temp/Extraction/Events/Run_1/Event_2
2) /path/to/temp/Extraction/Events/Run_1/Event_3
3) /path/to/temp/Extraction/Events/Run_1/Event_4
4) /path/to/temp/Extraction/Events/Run_1/Event_5
5) /path/to/temp/Extraction/Events/Run_1/Event_6
6) /path/to/temp/Extraction/Events/Run_1/Event_7
7) /path/to/temp/Extraction/Events/Run_1/Event_8
8) /path/to/temp/Extraction/Events/Run_1/Event_9
9) /path/to/temp/Extraction/Events/Run_1/Event_10

Enter range # of desired events (e.g. 1-5):
3-5 # User input                                                 
[####################] 100% Processing Event_6
All events processed!

Total Execution Time: 9626 ms
OBJ files written to: /tmp/EventOutput/MyData

Press ENTER to continue and upload files to the VR headset
\end{lstlisting}
% \begin{figure}[hb]
%     \centering
%     \includegraphics[width=1.0\linewidth]{Figures/CEDVRTerminal.png}
%     \caption{CEDVR Companion terminal interface. Upon pressing the enter key, the application uploads the OBJ files to the headset. }
%     \label{fig:ced_vr_terminal}
% \end{figure}

\section{Features}

\subsection{CEDVR Primary View}
\label{primaryscene}

A screenshot of the primary view of CEDVR is shown in Figure~\ref{fig:DetectorRefImage}.
The primary scene is in the frame of reference of the detector, in which the reconstructed objects are superimposed on a translucent detector with outlines. 
This provides a reference point with respect to the detector without obstructing the user's view. While the default settings display only outlines of the ECAL barrel, users also have the option to display outlines of other CMS detector elements. With unrestricted spatial movement and interactive zoom enabled by VR, this view allows users to fully immerse themselves in the three-dimensional structure of the event.

\subsection{CEDVR LEGO View}
\label{lego}

In proton collider physics, $p_T$, $\eta$, and $\phi$ are convenient variables for describing the momentum vector of reconstructed particles, while transverse energy $E_T$ is commonly used for calorimeter energy deposits and the vectorial sum known as missing-$E_T$~\cite{CMS_Experiment_2008}.
CEDVR therefore provides a LEGO visualization in $(\eta, \phi)$ space, displaying ECAL and HCAL $E_T$ deposits as well as an overlay showing the direction and transverse energy or momentum of reconstructed physics objects (jets, muons, electrons, and missing-$E_T$).

\begin{figure}[ht!]\centering \includegraphics[width=0.48 \textwidth]{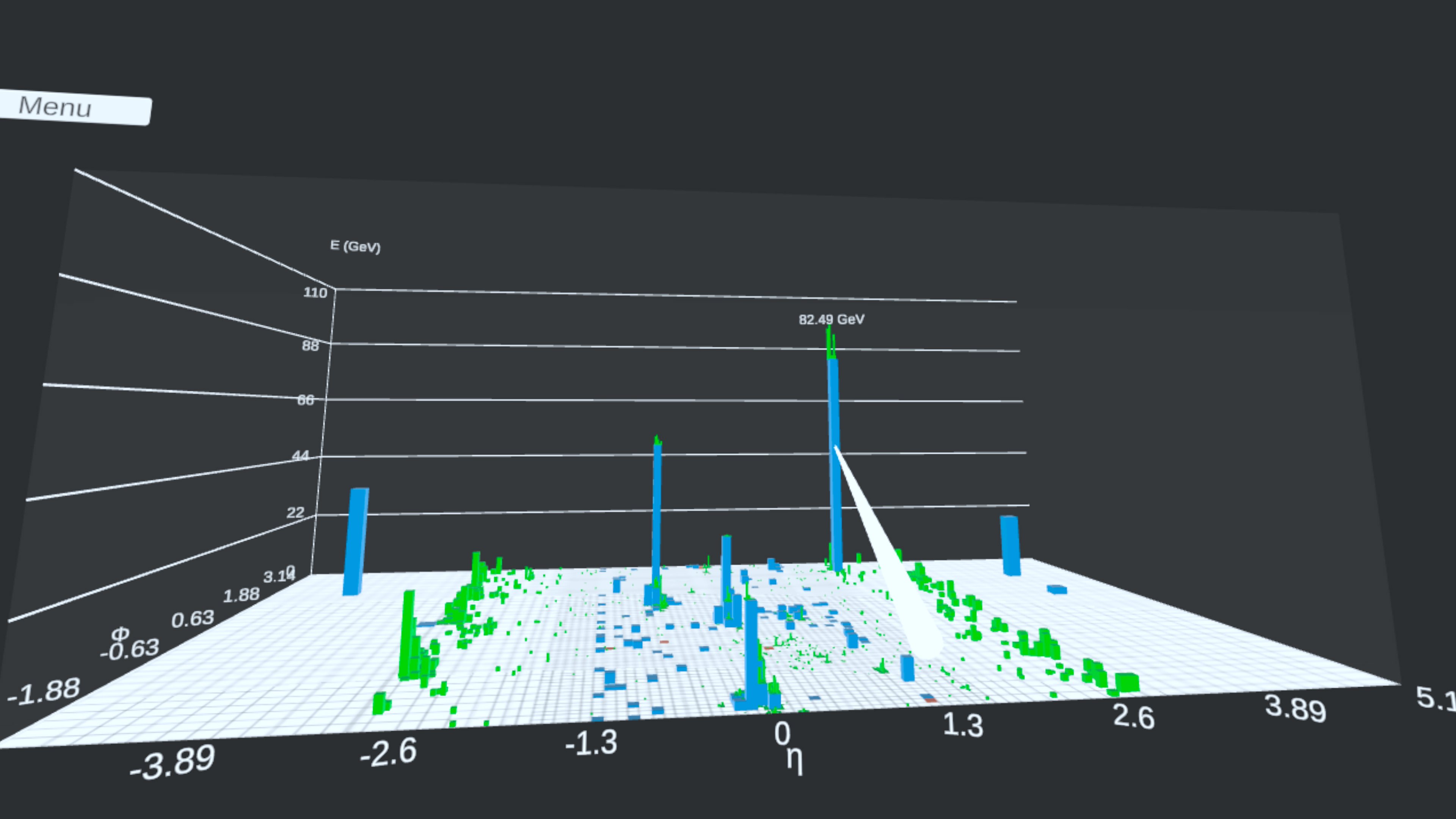}
\caption{ECAL and HCAL energy deposits depicted as towers in $(\eta, \phi)$ space in the CEDVR LEGO view, using an event from CMS Open Data \cite{CMSOpenDataBJetPlusXRun2012C}.}
\label{fig:LegoCalorimetry}
\end{figure}

\begin{figure}[ht!]\centering \includegraphics[width=0.48 \textwidth]{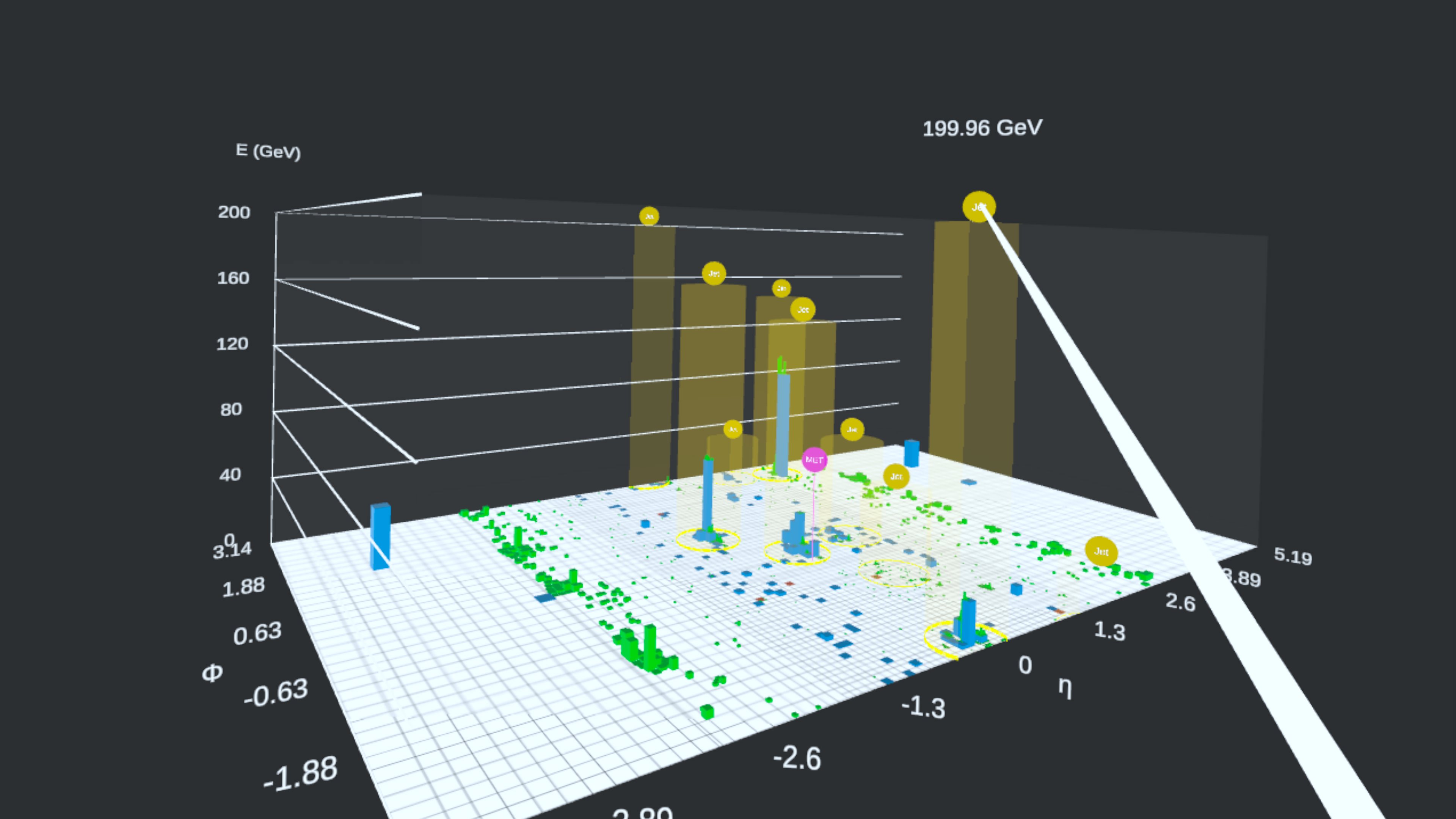}
\caption{A jet depicted as a cylinder in $(\eta, \phi)$ space in the CEDVR LEGO view, using an event from CMS Open Data
\cite{CMSOpenDataBJetPlusXRun2012C}.}
\label{fig:LegoJets}
\end{figure}

\begin{figure}[ht!]\centering \includegraphics[width=0.48 \textwidth]{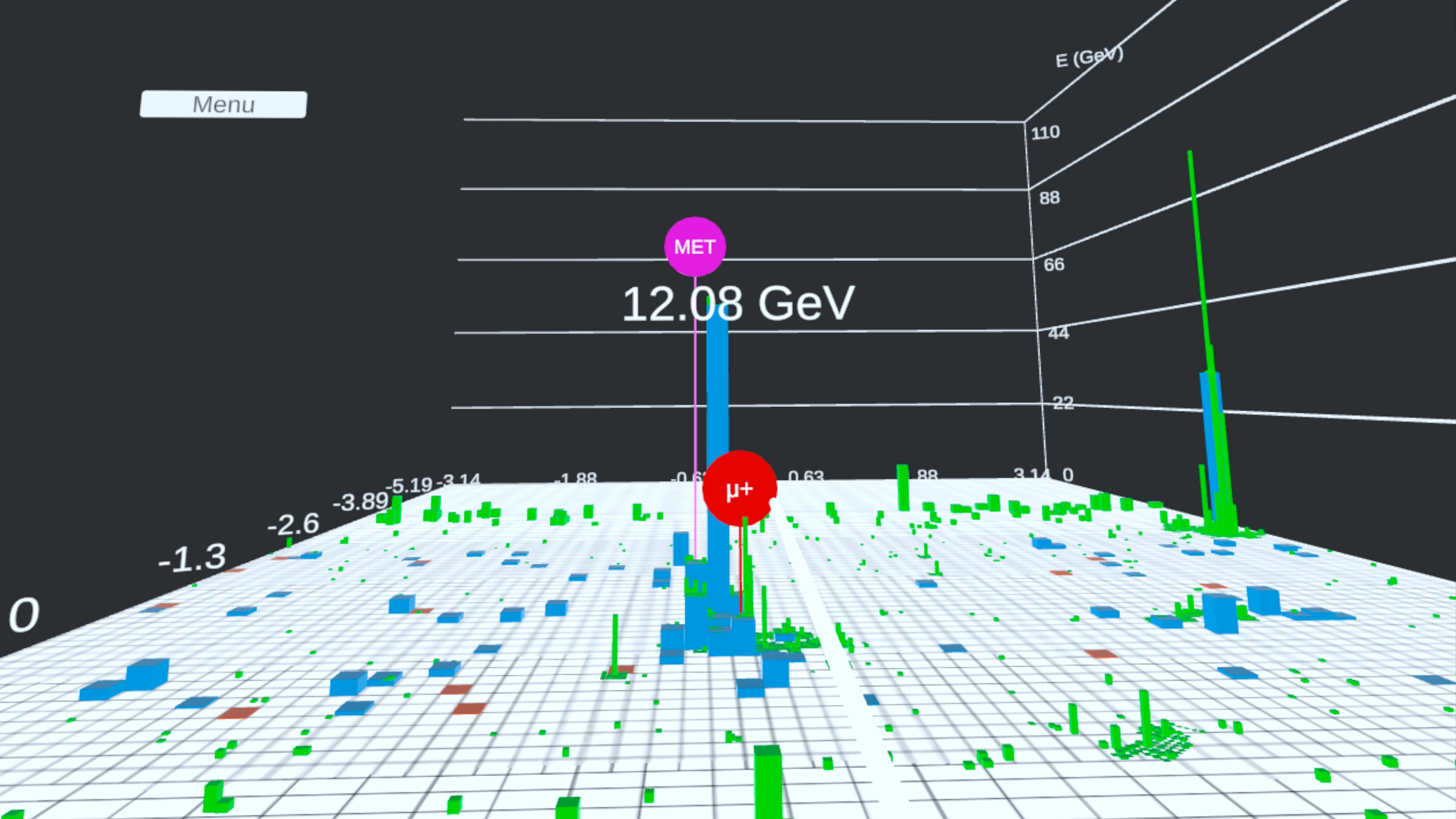}
\caption{A muon track depicted as a spike in $(\eta, \phi)$ space in the CEDVR LEGO view, using an event from CMS Open Data \cite{CMSOpenDataBJetPlusXRun2012C}.}
\label{fig:LegoMuon}
\end{figure}

Some screen captures from CEDVR's LEGO view can be seen in Figures~\ref{fig:LegoCalorimetry}--\ref{fig:LegoMuon}. In all of these figures, the tapering white cone is a handset pointer that is highlighting one particular calorimeter tower or physics object and causing its transverse energy or momentum to be displayed.

Figure~\ref{fig:LegoCalorimetry} shows $E_T$ deposits in different towers of the ECAL and HCAL.
Figure~\ref{fig:LegoJets} also shows a jet projected onto $(\eta, \phi)$ space, where it takes the form of a cylinder with a fixed height defined by its transverse energy. Notably, one of the jets in Figure~\ref{fig:LegoJets} crosses the boundary at $\phi=-\pi$. Since $\phi$ is periodic, part of the jet cylinder wraps around and is therefore displayed on the $\phi=\pi$ side as well. 
Figure~\ref{fig:LegoMuon} shows a positive muon ($\mu^+$) track projected onto $(\eta, \phi)$ space, where it takes the form of a sharp spike with a fixed height defined by its transverse energy. Since the missing-$E_T$ vector contains only transverse components, its $\phi$ position is always plotted at $\eta=0$.

The $(\eta,\phi)$ bins of the LEGO view are based on the segmentation of HCAL towers in the CMS detector. The HCAL towers are evenly segmented in the $(\eta,\phi)$ plane for $|\eta| < 1.740$. Beyond this point, the tower dimensions vary with $|\eta|$, producing a nonuniform LEGO grid. The detailed segmentation of the CMS HCAL can be obtained from the CMS Technical Design Report Volume 1 \cite{CMS-TDR-2006-001}. 
Since the fine ECAL grid in $(\eta,\phi)$ cleanly maps onto the coarser HCAL grid in most regions of the detector, ECAL energy deposits are stacked on top of HCAL energy deposits in the LEGO view. Similarly, HCAL depth segments (e.g.\ HB-1, HB-2, HE-1, HE-2) are also drawn as stacked layers within the same $(\eta,\phi)$ tower.

By disabling physics objects that are not of interest, users can customize each LEGO view according to their needs. To accommodate this flexibility, the vertical axis is automatically scaled to fit the highest-$E_T$ physics object in the scene. Any physics-based selection cuts that were applied to the regular event display also apply to the LEGO view.

\subsection{Menus and Object Inspection}
\label{menus}

\begin{figure}[t]
    \centering
    \includegraphics[width=0.95\linewidth]{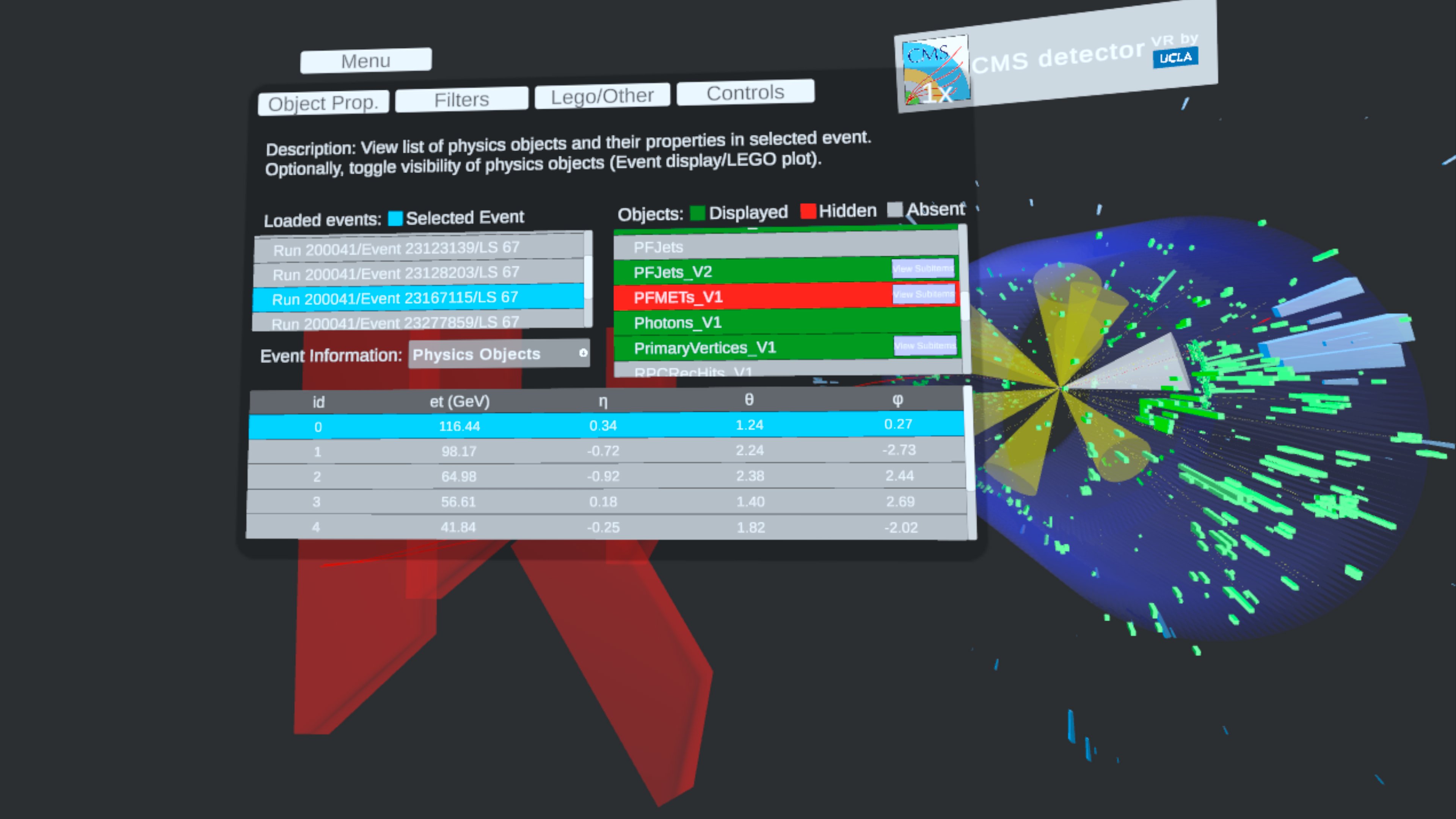}
    \caption{Menu with Object Properties tab selected, with a table showing which event has been selected, which objects are displayed, and an $E_T$-ordered list of physics objects, using an event from CMS Open Data \cite{CMSOpenDataJetHTRun2012C}.}
    \label{fig:ObjectPropMenu}
\end{figure}

\begin{figure}[t]
    \centering
    \includegraphics[width=0.95\linewidth]{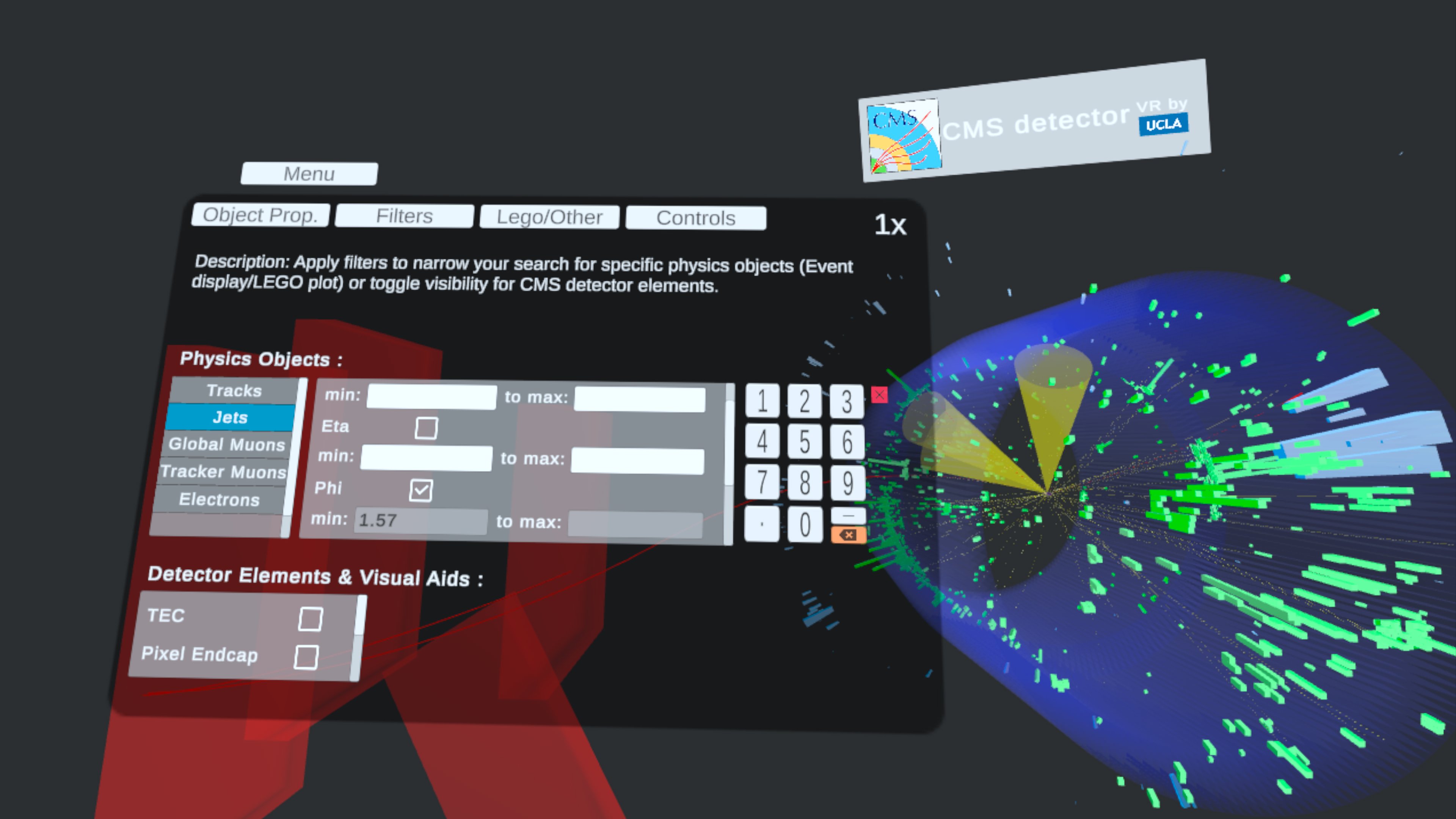}
    \caption{Menu with Filters tab selected, shown with a Jet filter that displays only jets with $\phi>1.57$, using
    the same event from CMS Open Data as in Figure~\ref{fig:ObjectPropMenu}.}
    %an event from CMS Open Data \cite{CMSOpenDataJetHTRun2012C}.}
    \label{fig:FilterMenu}
\end{figure}

CEDVR contains a single menu whose visibility and opacity can be adjusted. This menu contains four tabs: 
\begin{enumerate}
    \item Object Properties: This tab allows the user to cycle through events, toggle object visibility, as well as view sub-item data tables (Figure~\ref{fig:ObjectPropMenu}).
    \item Filters: This tab allows the user to apply selection requirements to the physics objects, as well as toggle detector elements and visual aids (Figure~\ref{fig:FilterMenu}).
    \item Lego/Other: This tab allows the user to view animations of the collision and access a LEGO view of the event.
    \item Controls: This tab contains a helpful diagram of controls for the application. 
\end{enumerate}

The event display contains several tools and features useful for analysis work. In the Object Properties tab, users can scroll through physics object collections in the event and view their individual properties using the sub-item tables which contain key observables such as charge, momentum, pseudorapidity ($\eta$), azimuthal angle ($\phi$), and transverse energy or momentum. They can also view an ``Energy Leaderboard" table which displays the most energetic objects in the event, and is particularly useful for analyses involving high-$p_T$ (transverse momentum) objects. Within these tables, users can also select specific physics objects of interest, which are then automatically highlighted in the scene for easy identification. Lastly, through the Filters tab of the menu, users can apply object selection cuts on tracks, jets, muons, and electrons across parameters such as charge, $p_T$, $E_T$, $\eta$, and $\phi$, with all specified cuts applied simultaneously.

\subsection{Additional Interaction Features}
\label{morefeatures}

\begin{figure}[ht!]
    \centering
    \includegraphics[width=1\linewidth]{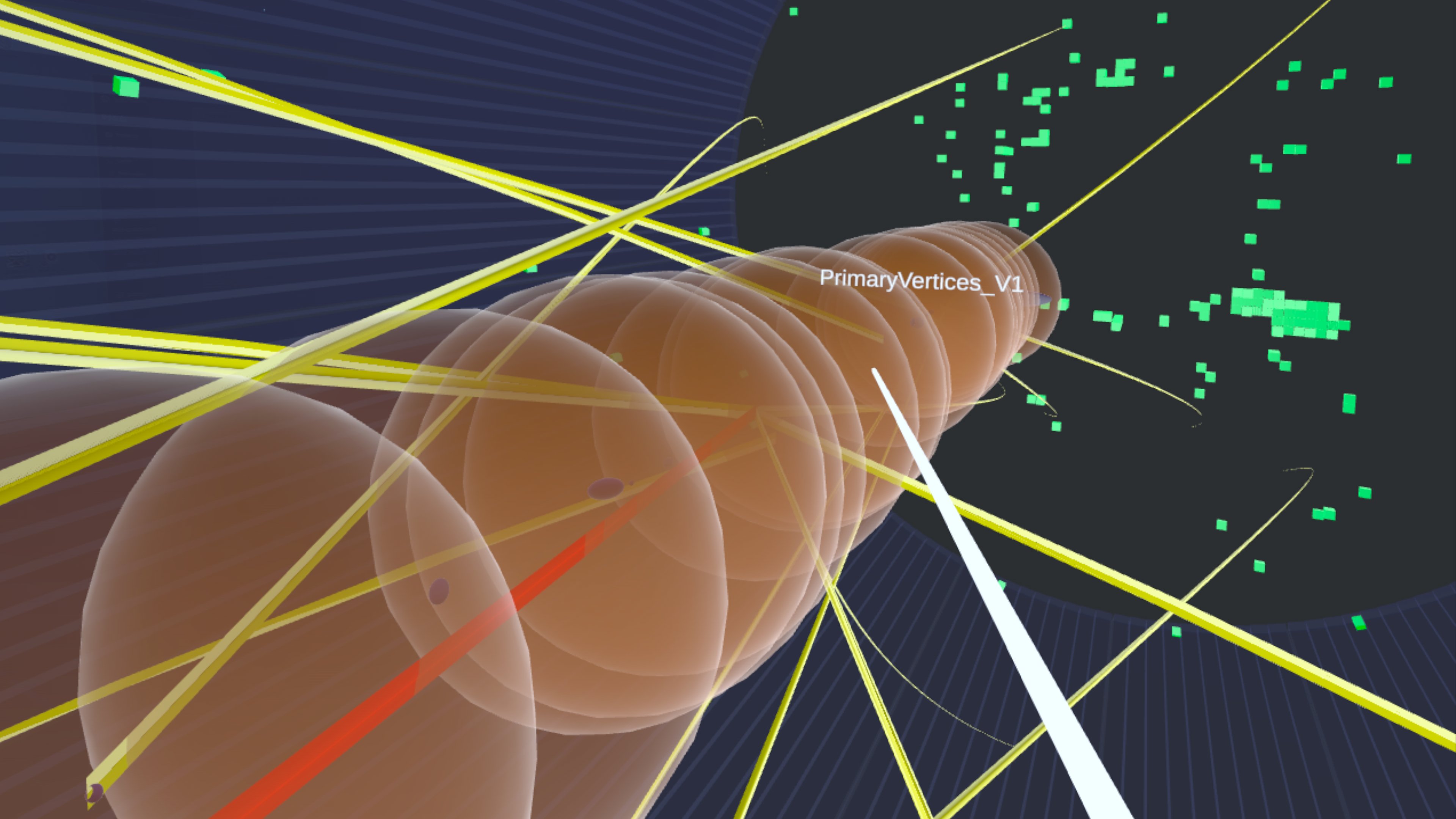}
    \caption{A close-up view of primary vertices (surrounded by locating spheres) using the zooming feature. Image produced from CMS Open Data
    \cite{CMSOpenDataJetHTRun2012C}.}
    \label{fig:PrimaryVertices}
\end{figure}

Additionally, there are several miscellaneous tools and capabilities useful for both analysis and outreach. Some key features are listed below:

\begin{itemize}
    \item Animation: Users can play an animation of the collision event and view tracks/muons propagating outward at constant display speed toward triggered detector elements. Since most particles produced in LHC collisions are highly relativistic and travel at nearly the speed of light, the animation enhances the visualization for educational outreach; it is not intended as a precise time-of-flight simulation.
    \item Scaled Ruler Axes: Users can toggle XYZ axes labeled with lengths to better gauge spatial scale and distances within the detector.
    \item Dynamic Track Resizing: Tracks taper and enlarge at the boundaries of the barrel to enhance visibility. This also qualitatively indicates the increase in spatial resolution outside of the detector's central region.
    \item Head-oriented zooming: Users are able to dynamically zoom in the direction that they are looking. This feature is particularly helpful when looking at small but detailed physics objects such as primary and displaced vertices, which are hard to see on a screen-based event display. This is illustrated in Figure~\ref{fig:PrimaryVertices}, where the primary vertices are indicated, along with the uncertainty in their reconstructed position, by the small blue ellipsoids, and locating spheres are added to help spot the vertices.
    \item Data Type Labels: Hovering over objects with a ray pointer produces a pop-up label with the name of the specific data type. This allows for rapid object identification in complex scenes, supporting accessibility in outreach efforts and analysis for researchers. 
\end{itemize}

\section{Performance}
\label{performance}

We conducted a performance evaluation of CEDVR on the Meta Quest 3 headset across seven events with varying levels of complexity. Here, we characterize event complexity by the total number of rendered physics objects, including tracks, jets, and calorimeter energy deposits. Object composition was consistent across all events, with calorimeter deposits comprising the majority of objects. One notable exception was the highest-complexity event, containing approximately 7,000 objects, the majority of which were tracks.

These measurements were performed using the OVR Metrics Tool, a performance-monitoring application developed by Meta. Each measurement followed a standard series of operations, including toggling the visibility of objects and rendering the LEGO view, which are computationally demanding operations. These operations were performed with the user stationary and were interspersed with fixed periods of idling.

Frame rate was averaged over the full duration of each measurement, starting after the loading of the scene. Using this measure, we find that, for typical events containing up to approximately 2,000 objects, CEDVR maintains stable performance exceeding 60 frames per second (FPS), providing smooth visualization. With more complex higher-multiplicity events, frame rate decreases. For instance, events with approximately 4,000 and 7,000 objects exhibit frame rates of 55 and 25 frames per second, respectively.

\section{Availability and System Requirements}
\label{availability}

CEDVR supports the Meta Quest 2/3/3S headsets and can be installed through a Meta release channel using the link provided on the CEDVR website~\cite{CEDVRwebsite}. No additional software is required to visualize the example events.

CMS scientists or open data users who wish to visualize their own selected CMS events need to prepare the data for visualization in the headset. First, as described in Section~\ref{analyzer}, users must run the iSpy Analyzer code on CMS data to create ig files, which are zip archives containing event information in JSON format. In order to properly visualize calorimetry data in CEDVR's LEGO view, users must also modify the iSpy Analyzer according to the instructions provided in the CEDVR Companion Parser's GitHub README, accessible through our website (\url{https://vr.physics.ucla.edu/cedvr.html}). Then, as described in Section~\ref{parser}, the user must run the CEDVR Companion Parser app on a laptop or desktop computer to convert the ig file into OBJ and JSON files and upload them to the VR headset. The CEDVR Companion Parser supports macOS/Linux/Windows, and is available as a command-line executable via GitHub. The CEDVR Companion Parser uses the Android Debug Bridge (ADB) to transfer processed event files to the headset. ADB is included in Google's freely available Android SDK Platform-Tools package and can be installed on the laptop or desktop computer independently of Android Studio. The Quest headset must have developer mode and USB debugging enabled before it can communicate with ADB over a USB connection.

\section{Conclusions}
\label{conclusions}

In this work, we have presented CEDVR, a virtual reality-based event display toolkit for the CMS experiment designed to complement existing displays with an immersive three-dimensional perspective. Designed with a focus on physics analysis and ease of use, CEDVR maintains a simple menu structure that provides functional tools for filtering and visualizing data, including object-level information tables and access to the LEGO view. Integrated within existing CMS workflows, CEDVR and the CEDVR Companion Parser allow physicists to easily load event data and quickly visualize reconstructed CMS events in virtual reality. On a Quest 3 headset, CEDVR maintained frame rates above 60 FPS for typical events containing up to approximately 2000 rendered objects. Its simple controls and the inclusion of several demonstration events make CEDVR well suited for education and public outreach. CEDVR has been showcased extensively within the CMS Collaboration, and at public outreach events. 

Looking ahead, continued advances in consumer VR hardware are expected to enable further enhancements in display resolution and rendering performance. 
Future development of CEDVR is expected to add additional visualization and analysis capabilities.
The Phase 2 upgrade of the CMS detector for the High-Luminosity LHC will significantly increase event complexity through higher pileup conditions and new detector subsystems, including upgraded trackers, high-granularity calorimetry, the addition of timing detectors with enhanced precision, and an extended muon system \cite{Pasztor:2022CMSPhase2}. These changes will necessitate the update of CEDVR's detector geometry and data pipeline architecture as well as performance optimizations to accommodate larger event sizes. CEDVR's modular design should allow it to adapt to evolving CMS data formats and support VR event visualization for the Phase-2 CMS detector.

\section*{Acknowledgements}
\label{ack}

We would like to acknowledge and thank Thomas McCauley and the rest of the iSpy developers for their work on the iSpy analyzers – which act as input for CEDVR – and for valuable feedback on our work. This work was supported by U.S. CMS Operations at the LHC under NSF grant 2121686 and the U.S. Department of Energy under award number DE-SC0009937.

\bibliographystyle{elsarticle-num}

\bibliography{CEDVRbib}

\end{document}